# Molecular anchoring stabilizes low valence Ni(I)TPP on copper against thermally induced chemical changes

Henning Maximilian Sturmeit,[1] Iulia Cojocariu,[2] Matteo Jugovac,[2,#] Albano Cossaro,[3] Alberto Verdini,[3] Luca Floreano,[3] Alessandro Sala,[3,4] Giovanni Comelli,[3,4] Stefania Moro,[4] Matus Stredansky,[3,4] Manuel Corva,[3,4] Erik Vesselli,[3,4] Peter Puschnig,[5] Claus Michael Schneider[2,6], Vitaliy Feyer,[2,6,*] Giovanni Zamborlini,[1,*] and Mirko Cinchetti[1].

[1]Technische Universität Dortmund, Experimentelle Physik VI, 44227 Dortmund, Germany.

[2]Peter Grünberg Institute (PGI-6), Forschungszentrum Jülich GmbH, Jülich, Germany

[3] CNR-IOM, Lab. TASC, s.s. 14 km 163,5, 34149 Trieste, Italy

[4] Physics Department, University of Trieste, 34127 Trieste, Italy

[5]Institut für Physik, Karl-Franzens-Universität Graz, NAWI Graz, 8010 Graz, Austria

[6] Fakultät f. Physik and Center for Nanointegration Duisburg-Essen (CENIDE), Universität Duisburg-Essen, 47048 Duisburg, Germany

[#]Present address: Istituto di Struttura della Materia-CNR (ISM-CNR), Trieste, 34149, Italy

*Corresponding authors: v.feyer@fz-juelich.de, giovanni.zamborlini@tu-dortmund.de.

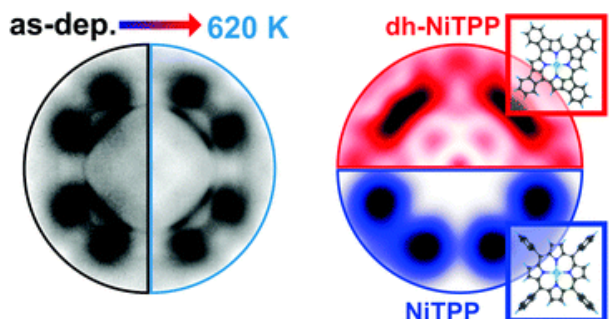

## Abstract

Many applications of molecular layers deposited on metal surfaces, ranging from single-atom catalysis to on-surface magnetochemistry and biosensing, rely on the use of thermal cycles to regenerate the pristine properties of the system. Thus, understanding the microscopic origin behind the thermal stability of organic/metal interfaces is fundamental for engineering reliable organic-based devices. Here, we study nickel porphyrin molecules on a copper surface as an archetypal system containing a metal center whose oxidation state can be controlled through the interaction with the metal substrate. We demonstrate that the strong molecule–surface interaction, followed by charge transfer at the interface, plays a fundamental role in the thermal stability of the layer by rigidly anchoring the porphyrin to the substrate. Upon thermal treatment, the molecules undergo an irreversible transition at 420 K, which is associated with an increase of the charge transfer from the substrate, mostly localized on the phenyl substituents, and a downward tilting of the latters without any chemical modification.

## Introduction

Metalloporphyrins on metal substrates frequently self-assemble in 2-dimensional (2D) lattices,[1,2] where the chemical state and the adsorption properties of each molecule are well-defined and homogeneous across the whole surface, providing a unique platform for catalytic applications, gas-sensing, and spintronics. Metalloporphyrins are characterized by an ion center, which can act as an axial coordination site for anchoring ligands. Small molecules, such as common or toxic gasses, can be stabilized at the macrocycle moiety and become then available for a subsequent reaction, e.g., oxygen or nitric oxide reduction.[3] Moreover, if the binding of a ligand produces significant changes in the overall electronic properties of the porphyrin, the molecular layer becomes, to all intents and purposes, a gas sensor.[4] In the field of on-surface magnetochemistry,[5] these effects have been exploited to control exchange coupling between a molecular layer and a ferromagnetic substrate.[6,7] The added ligand competes with the surface underneath the porphyrin for the stronger bond at the metal center, giving rise to the so-called surface trans effect,[8–11] a very intriguing method to control the spin, electronic and chemical properties of metal/molecular interfaces in the field of molecular spintronics[12] and catalysis.

All these applications require thermally and chemically stable interfaces that can be regenerated at will.[13] Temperature annealing is one of the most effective ways to restore the molecular layer to its pristine configuration by removing anchored ligands that alter the activity of the catalytic center. In powders, porphyrin molecules are usually robust against thermal treatments as they can be annealed to high temperatures. However, when deposited on a surface, the molecule–substrate interaction drastically influences the stability of the molecular film. If the interaction is weak, the molecules may desorb from the substrate before the anchored ligand has been removed or, in the opposite case, the substrate can accelerate the thermal decomposition of the molecular layer. It was also shown that, for selected prototypical templates, e.g., silver, copper and gold, the on-surface annealing promotes the partial dehydrogenation of polycyclic aromatic hydrocarbons,[14,15] which may lead to the formation of organometallic oligomers[16,17] or to the cyclodehydrogenation of peripheral ethyl groups.[18–20] Furthermore, also intramolecular structural changes can occur at the macrocycle moiety within the porphyrin core.[21–25] These changes irreversibly affect the electronic structure of the molecule, as shown in the case of iron(II)-octaethylporphyrin deposited on a gold electrode, where the cyclodehydrogenation reaction alters the magnetic anisotropy of the Fe ion, thus irreversibly modifying its effective spin moment.[26]

The present work focuses on the thermal stability of Ni(I)-containing tetraphenyl porphyrin (NiTPP) molecules, self-assembled on the copper (100) surface. Whereas in the gas-phase the nickel ion is in the formal 2+ oxidation state (Ni(II)), the charge transfer taking place upon the adsorption of NiTPP on the copper surface leads to a stabilization of the Ni(I) species[27,28] and to a partial filling of the former lowest unoccupied molecular orbitals (LUMOs) up to the LUMO+3.[29] The uncommon Ni(I) oxidation state can also be found in the Ni-containing porphyrinoid core of the biological coenzyme F430.[27] This coenzyme plays a key role in both the methanogenesis and the anaerobic methane oxidation and owes its reactivity specifically to the 1+ metal oxidation state.[30]

We use a variety of surface-sensitive techniques to demonstrate the thermal and chemical stability of NiTPP layer on Cu(100), together with its reactive Ni(I) center, up to an annealing temperature of 620 K. By means of scanning tunneling microscopy (STM), we detect the occurrence of a thermally activated transformation that, as verified by X-ray photoemission spectroscopy (XPS), starts already at 360 K. From the topographic STM images, we observe an apparent height decrease of the NiTPP phenyl terminations and at the same time an overall shift of their C 1s related component towards lower binding energy. Photoemission tomography (PT), infrared-visible sum-frequency generation (IR-Vis SFG) spectroscopy and near-edge absorption fine structure spectroscopy (NEXAFS) exclude chemical changes at the macrocycle center. No dehydrogenation followed by either a ring-closing reaction or a C–Cu bond formation takes place in the molecular film, in contrast with similar systems reported in the literature.[21-24,31] In fact, the strong interaction between the as-deposited NiTPP film and the copper substrate, which is accompanied by an elevated charge transfer from the copper substrate to the NiTPP, is responsible for anchoring the molecule to the substrate in its long-range ordered monolayer phase. This prevents any molecular displacement that would allow a true flattening of the phenyls and favor their cyclodehydrogenation. The transformation is rather associated with an increase of charge transfer from the substrate to the molecule, mostly localized on the phenyl substituents that undergo a small downward bending without further chemical modifications. The Ni(I) oxidation state of the NiTPP molecule is thus preserved in the thermally treated molecular layer.

The stability of this molecular array is a key factor for employing this system effectively in single-atom catalysis, biosensing, and on-surface magnetochemistry applications, where thermal cycles are commonly used for regenerating the pristine properties of the active molecular layer.

## Methods

### Sample preparation

The clean Cu(100) surface was prepared by a standard procedure: cycles of Ar$^+$ ion sputtering at 2.0 keV followed by annealing to 800 K. NiTPP molecules (Sigma Aldrich, 95% purity) were thermally sublimated at 570 K from a Knudsen cell type evaporator onto the copper substrate kept at room temperature. In the STM and IR-Vis SFG set-ups, as well as at the ALOISA end station, the evaporation rate was checked by means of a quartz micro-balance. At the NanoESCA beamline, instead, the NiTPP deposition was optimized by LEED. Close to the 1 saturated ML coverage regime, sharp LEED patterns were observed, while additional deposition on an already saturated surface blurred the diffraction spots.

## X-ray photoemission and absorption spectroscopies

The valence band spectra and the momentum maps were measured at the NanoESCA beamline of Elettra, the Italian synchrotron radiation facility in Trieste, using an UHV (base pressure better than $5 \times 10^{-11}$ mbar) electrostatic photoemission electron microscope (PEEM) set-up described in detail in ref. [32]. In our PEEM set-up, two-dimensional momentum maps are collected in a single-shot experiment, with a wide reciprocal space range: $k_x, k_y \in [-2.1; +2.1]$ Å$^{-1}$. By means of a double hemispherical electron analyzer, it is possible to select a specific kinetic energy of the photoemitted electrons, thus measuring the full energy *vs.* parallel momentum data cube. The data were collected with a photon energy of 35 eV and a total energy resolution of 70 meV, using linearly p-polarized light, while keeping the sample at 90 K during the measurements. The NEXAFS and X-ray photoemission experiments were performed at the ALOISA beamline, also located at Elettra synchrotron.[33] The spectra across the C and N K-edges were acquired in electron yield mode using a channeltron multiplier,[33] and analyzed following the procedure described in ref. [34]. The orientation of the surface with respect to the linear polarization (s and p) of the synchrotron beam was changed by rotating the sample around the beam axis while keeping fixed the incident angle (6° with respect to the surface plane) of the synchrotron light. The core level spectra were collected in normal emission geometry with a total energy resolution (analyzer + beamline) of 300 meV. In the C 1s fit of the NiTPP/Cu(100) upon annealing, the ratios between the areas corresponding to the inequivalent carbon species were kept fixed to 24 : 4 : 8 : 8. These ratios correspond to the numbers of phenyl, *meso*-bridge, C–C–N pyrrolic and C–C–C pyrrolic carbon atoms within the NiTPP, respectively. The resulting Lorentzian width was 690 meV for all components, while the experimental resolution and inhomogeneity contributions were accounted for by a Gaussian broadening of 300 meV, as obtained from the fit. The BE values for the four C species of the pristine NiTPP were kept fixed during the fit procedure of the C 1s spectra measured as a function of the annealing temperature. The binding energy of the flatter phenyl component was determined from the fitting of the spectrum annealed at the highest temperature (470 K), while keeping all the other parameters for the other C species fixed.

## STM

Low-temperature STM experiments were carried out with an Omicron LT-STM system located in the TASC laboratories, Trieste, Italy. All the measurements were performed at a sample temperature of 77 K. The microscope is hosted in an ultrahigh vacuum (UHV) chamber, operating at a base pressure below $7 \times 10^{-11}$ mbar. Images were acquired in constant current mode, with inverse bias voltage applied to the tip while the sample was grounded. Electrochemically etched tungsten tips were used for imaging. At each annealing step, the sample was kept at the corresponding temperature for 5 min and then cooled down to 77 K in the STM before being measured.

## IR-Vis SFG

The data were collected in a specifically dedicated experimental setup, described elsewhere,[35] at the Physics Department of the University of Trieste. In brief, an ultra-high vacuum (UHV) system with a base pressure of 5 × 10$^{-11}$ mbar, equipped with standard surface science preparation and characterization techniques, is directly attached to a variable pressure reaction cell for *in situ* IR-Vis SFG measurements. The sample can be transferred directly from the preparation chamber to the reaction cell and *vice versa* without breaking the vacuum. The laser excitation source (Ekspla) delivers a 532 nm (2.33 eV, 30 ps pulses at 50 Hz) visible beam and tunable IR radiation in the 1000–4500 cm$^{-1}$ range. The raw SFG spectra were normalized by the non-resonant signal generated by a reference GaAs crystal in order to account for modulations in the IR intensity, both intrinsic and due to gas phase adsorption along the optical path in air before entering the cell. Data were then analyzed by least-squares fitting to a parametric, effective expression of the nonlinear second-order susceptibility:[35–37]

$$\frac{I_{SFG}(\omega_{IR})}{I_{vis}I_{IR}(\omega_{IR})} \propto \left| A_{NRes} + \sum_k \frac{A_k e^{i\Delta\varphi_k}}{\omega_{IR} - \omega_k + i\Gamma_k} \right|^2$$

The model lineshape accounts for the amplitudes of the nonresonant ($A_{NRes}$) and *k*th resonant ($A_k$) contributions and includes the phase difference between each resonance and the nonresonant signal ($\Delta\varphi_k$), the position of the resonances ($\omega_k$), and their Lorentzian broadening ($\Gamma_k$). The latter is related to the dephasing rate of the excited state, originating from the lifetime and the elastic dephasing of the excited vibronic state.[35,38] In the manuscript's figures, we plot the normalized IR-Vis SFG signal (grey dots) together with the best fit (black lines). For better visual clarity, we also plot (filled profiles) the intensity of the resonances and their interference with the nonresonant background, calculated with the parameters obtained from the fitting procedure, according to the following expression:[35,37]

$$I_{SFG,k}(\omega_{IR}) \sim \left| A_{NRes} + \frac{A_k e^{i\Delta\varphi_k}}{\omega_{IR} - \omega_k + i\Gamma_k} \right|^2$$

We recall that the fitting procedure provides an optimal set of lineshape parameters that does not represent a unique solution, since several (2*N* − 1 for *N* resonances) equivalent sets exist.[39] This originates from the fact that only intensities are measured. However, as long as we are interested in amplitudes and variations of relative phases, this point does not represent an issue in our case. The lineshape parameters of the resonances, as obtained from the optimization procedure, are reported in Table S1 (ESI†).

### Density functional theory

We have performed density functional calculations for gas-phase NiTPP ($NiC_{44}N_4H_{28}$) and for a possible de-hydrogenation product after annealing dh-NiTPP ($NiC_{44}N_4H_{20}$) using the NWCHEM code[40] employing the B3LYP hybrid functional for exchange–correlation effects.[41,42] The simulated momentum maps of the gas-phase molecules were obtained as the Fourier transforms of the respective Kohn–Sham orbitals as described previously.[43,44] Further details about the DFT simulations and a comparison of the molecular orbitals of NiTPP and dh-NiTPP can be found in the ESI.†

### Results

STM is a widely employed technique for studying the topography and electronic structure of molecular arrays supported by metal electrodes. Here, constant-current STM images of the NiTPP/Cu(100) interface were acquired at 77 K, starting from the as-deposited layer (room temperature) and after annealing at temperatures of 420 and 530 K (further details about the sample annealing can be found in the Methods section). The resulting images are shown in Fig. 1a. The STM appearance of the as-deposited (as-dep.) NiTPP molecule is dominated by four bright features and a dark depression in the center, associated with the four peripheral phenyl rings and with the macrocycle core, respectively. As already shown in ref. [29], this counterintuitive appearance originates from the strong molecule–surface interaction. In fact, the macrocycle lies very close to the copper termination (2 Å), forcing the phenyl rings to point away from the surface, as revealed by previous DFT calculations.[29] A ball-and-stick model of the gas-phase NiTPP molecule, indicating the different molecular moieties, is reported in Fig. 1b.

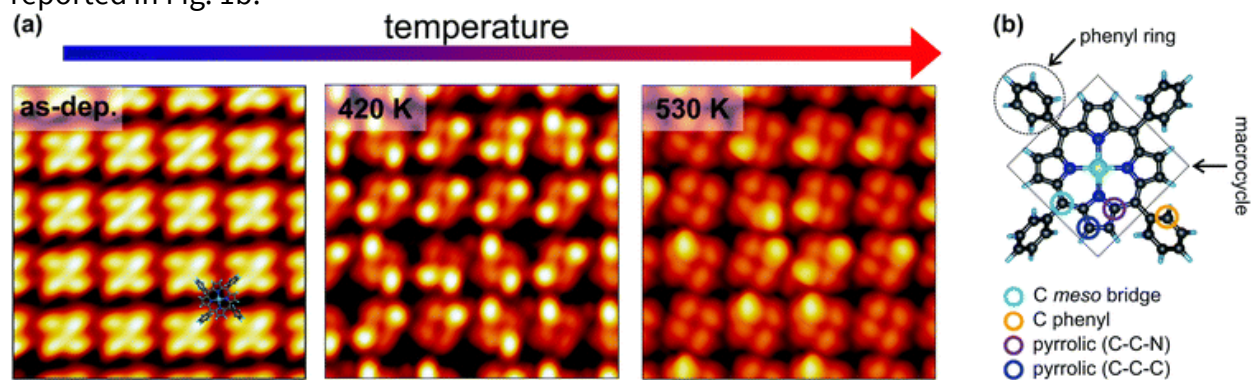

**Fig. 1** Temperature-dependent topographic STM images of the NiTPP film on Cu(100). (a) STM image parameters: lateral size 5.0 × 5.0 nm$^2$, tunneling parameters: left −1.5 V/50 pA; center −0.8 V/100 pA; right −0.5 V/500 pA. A sketch of the NiTPP molecule is superimposed to the STM image of the as-deposited molecular film (left). The appearance of a single NiTPP in the molecular lattice was identified in ref. [29]. (b) Ball-and-stick model of the gas-phase NiTPP molecule. Each atomic species is painted with a different color: Ni in green, N in blue, C in dark grey and H in light blue. The different carbon species and molecular moieties are labeled.

Already after the first annealing step, the apparent height of some of the phenyl lobes decreases by several tenths of Å, and after annealing to 530 K, almost all the lobes display a dimmed contrast. The contrast polarity is independent of the bias and the tunneling current. A statistical analysis based on large-scale STM images revealed that, after annealing to 420 K, the height change involves 51 ± 7% of the phenyls, while at 530 K the percentage increases to 87 ± 9%. We did not notice any specific pattern in the propagation of phenyl flattening, *i.e.*, random combinations of pristine bright and annealed dim phenyls within the molecule are observed in the STM images taken at different temperatures, ranging from all four phenyls in the as-deposited geometry to all four phenyls in the final configuration. We also observed that the dimming of one specific phenyl ring does not systematically influence the intensity/contrast of neighboring NiTPP molecules. From a comparison with former reports on silver[22,25,45] and copper[21,23,46,47] surfaces, one might tentatively associate the apparent decrease of the phenyl height with a molecular flattening due to partial cyclodehydrogenation and fusion of the phenyls with the tetrapyrrolic aromatic moiety. However, a completely flattening of the phenyls is not fully compatible with the topographic images on the basis of simple steric arguments, because we do not observe any change in the intermolecular distance or the positional alignment with respect to the substrate, whereas flattened TPPs would require a larger footprint area in either flat conformation (spiral, rectangular and hybrids as mapped on silver),[22,25,45] hence a change of surface phase density and symmetry.[48] In fact, not even the lattice phase symmetry is affected by the observed transformation and the NiTPP lattice can still be described using the (4,3/−3,4) and (3,4/−4,3) matrices reported in ref. [29], as confirmed by the LEED patterns taken on the as-deposited film and after annealing to 520 K (see Fig. 1S, ESI†). Overall, the preservation of the phase symmetry, as well as of the local molecular arrangement across the transformation, indicates that the porphyrin molecules are strongly anchored to the surface.

Nevertheless, at the present level of detail, the STM topography data cannot directly determine whether the transformation is associated with a chemical reaction, a conformational change or a variation of the density of states in the phenyl moiety. From a chemical point of view, the most important evidence is the absence of any changes in the energy position and lineshape of the Ni 2p core level after annealing at 470 K, as shown in the XPS spectra in Fig. 2a. The binding energy of the Ni 2p core level is 853.15 eV. As reported in ref. [27], this specific binding energy is a fingerprint of the Ni(I) oxidation state responsible for the catalytic activity of the metal porphyrin 2D layer. In general, XPS alone is not sufficient to properly determine the oxidation state of a metallic ion; however, our claims are further supported by the Ni $L_3$-edge NEXAFS spectra measured with s- and p-polarized light (s-pol and p-pol, respectively), shown in Fig. 7S (ESI†). The s-pol absorption spectrum shows only a single peak at 852.4 eV associated with the σ*-symmetry transition from the $2p_{3/2}$ to $d_{x^2-y^2}$ levels, while no strong satellites are present at higher photon energies. Such a lineshape is fully consistent with a low valence nickel ion (Ni(I)) configuration with partial unoccupied $d_{x^2-y^2}$ level.[27,49,50] The strong linear dichroism in the spectra, *i.e.* the difference between the spectra measured with s- and p-pol, is consistent with the expected σ-symmetry of the main peak corresponding to a closely planar macrocycle oriented parallel to the surface. The energy position of the resonances and the dichroic signal at the Ni $L_3$-edge are characteristic of a chelated Ni(I) ion rather than a Ni adatom or a NiCu alloy, excluding the replacement of the central Ni ion by a Cu atom[27] (also referred as transmetalation), which was previously reported for similar organic/metal interfaces.[51,52] The N 1s XPS spectrum also shows only a minimal shift to lower binding energy of less than −0.1 eV (see Fig. 2S, ESI†), which suggests either a small increase of the charge transfer from the substrate or a small decrease of the distance between the nitrogen atoms and the surface. The spectroscopic evidence thus indicates that the overall reduced oxidation state of the tetrapyrrolic pocket is not affected by the transformation at 420 K. These observations prove the exceptional robustness of the present system against high temperatures as a consequence of the strong molecule–substrate interaction, which is responsible for the charge injection on the macrocycle and its Ni center. As a consequence, we can confidently associate the local anchoring of the molecules to an effective interaction of the macrocycle moiety with the Cu(100) surface, which persists throughout the annealing process.

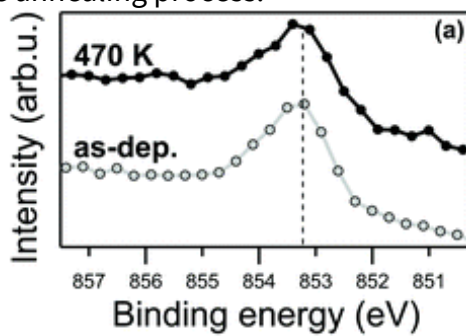

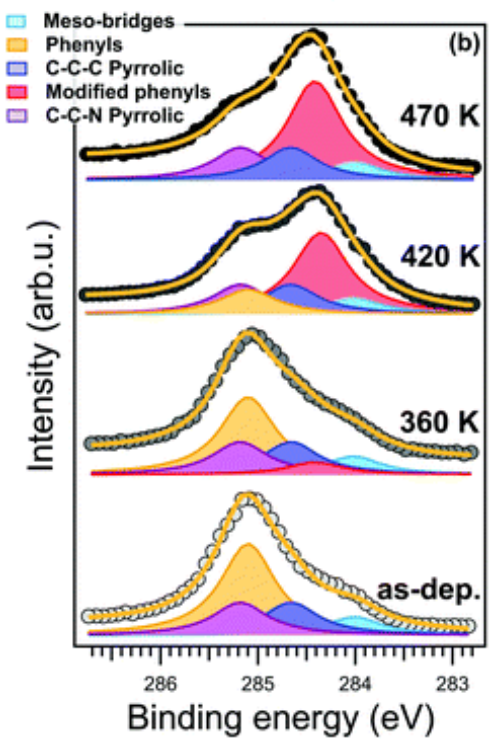

**Fig. 2** XPS spectra of NiTPP/Cu(100) interface at different core level shells. (a) Ni $2p_{3/2}$ spectra of as-deposited (bottom) and annealed up to 470 K (top) film. (b) C 1s spectra (data and fit) of the NiTPP/Cu(100) interface taken after annealing treatment to the indicated temperature. All spectra were collected in normal emission geometry with a photon energy of 1020 (a) and 515 (b) eV, respectively.

In contrast to the Ni 2p and N 1s, the C 1s lineshape clearly changes after the annealing. The C 1s core level spectra of the NiTPP/Cu(100) interface is reported in Fig. 2b. The C 1s spectrum of the as-deposited NiTPP monolayer at room temperature can be fitted using a multi-peak Voigt function, which accounts for the four different carbon species contained in the NiTPP molecule, as depicted in Fig. 1b(further details can be found in the Methods section). The 44 C atoms are subdivided in 24 phenylic (binding energy 285.1 ± 0.1 eV), 4 *meso*-bridges (284.0 ± 0.1 eV), 8 C–C–N pyrrolic (285.2 ± 0.1 eV), and 8 C–C–C pyrrolic (284.6 ± 0.1 eV) carbon atoms (see Fig. 2b). The overall energy alignment of the different components is in agreement with the results of a ZnTPP multilayer on Si(111) reported by Cudia *et al.*[53]

Already after annealing to about 360 K, a new component arises at 284.4 ± 0.1 eV binding energy (BE), see Fig. 2b. For higher annealing temperatures, it becomes clear that, while the area of the new component grows, the one associated with the phenyl moieties in the pristine layer decreases by the same amount, without appreciable changes of the total C 1s area, hence the new feature can be confidently associated with the fraction of phenyl carbons undergoing a transformation across the critical temperature transformation previously identified by STM. After annealing to 470 K, all of the carbon atoms of the pristine phenyl moiety have been transformed into the new state. The conservation of the area below the phenyl peaks suggests that the moiety remains intact on the surface after the temperature treatment, *i.e.*, no partial dehydrogenation may take place. At the same time, we can exclude the formation of new C–Cu bonds due to partial cyclodehydrogenation of the carbon atoms, which otherwise would give rise to a characteristic XPS component close to the binding energy of 283.2 eV, as was reported for other compounds.[31] Time-dependent C 1s XPS measurements show that the transformation is irreversible within our experimental time-scale (>24 hours).

The corresponding shift by ∼0.7 eV to lower binding energy appears to be too large for being attributed solely to a final state screening effect due to a smaller distance between the phenyls and the copper surface. Thus, these arguments do not allow us to discriminate between the case of a bare charge transfer to the phenyls or a more drastic chemical modification that would affect the aromatic structure of the whole molecular complex.

In order to shed light on this point, we exploited photoemission tomography (PT) to directly probe the electronic structure of the macrocycle. PT is an experimental approach that combines two-dimensional momentum mapping with density functional theory (DFT) calculations. Under specific assumptions about the final state of the photoemitted electron, the momentum pattern arising from a particular initial molecular state can be written as the Fourier transform (FT) of the initial-state wave function.[43] Thus, this relation provides a one-to-one correspondence between the angular distribution of the photocurrent and the molecular orbitals in the reciprocal space. In this way, molecular orbitals can be identified from their momentum pattern[54–57] or their wave functions can be retrieved using iterative computational procedures[58,59] (for further details see the Methods section).

In Fig. 3a, we compare the angle integrated valence band (VB) spectrum of the clean substrate and of the as-deposited NiTPP film with those measured after a stepwise annealing of the sample up to 720 K. The spectra were obtained by integrating the intensity in the whole available $k_\parallel(k_x,k_y)$ range as a function of the electron binding energy referred to the Fermi level

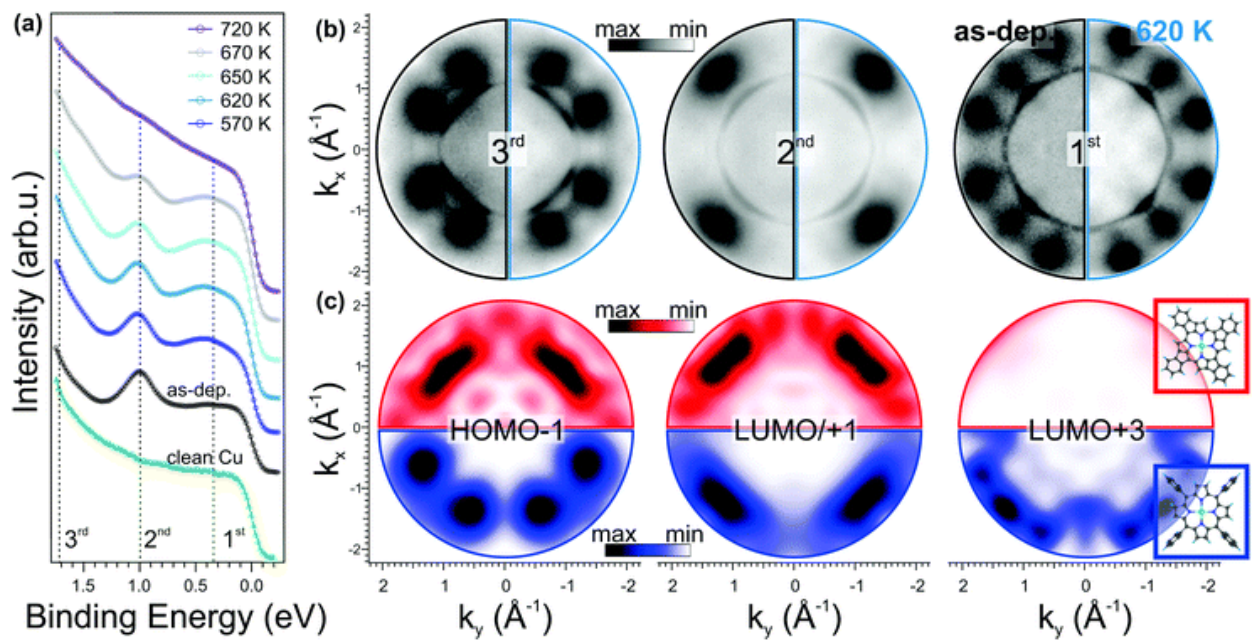

($E_F$) of the copper substrate. The clean copper has a rather featureless spectrum between 0 and 1.8 eV BE, while after 1 saturated ML NiTPP deposition, three new features (marked with dashed lines) appear in the spectrum. These features have a clear fingerprint in the momentum space, as will be discussed below. At first sight, the annealing treatment does not significantly affect the spectra: at a temperature of 570 K, we observed only a slight increase of the intensity of the VB region, in particular of the feature at 0.4 eV, while no shift in the energy position of the main molecular states is observed. Beyond 620 K, the VB spectrum starts to change. The two components at 0.4 and 1 eV gradually lose intensity, suggesting partial decomposition of the molecules in the layer that is completed at 720 K, where the molecular features are no longer visible. This provides a further indication of the thermal stability of the molecular film up to 620 K, in full agreement with the XPS C 1s spectra taken in the 620–720 K temperature range (see ESI,† Section 3).

**Fig. 3** Angle-integrated valence band and photoemission tomography data of NiTPP/Cu(100) interface. (a) Valence band spectra acquired after annealing at the indicated temperatures. (b) Momentum patterns of pristine (left) and annealed (right) films, BE indicated on (a). (c) Calculated FTs of the molecular orbitals of NiTPP (bottom) and dh-NiTPP (top).

In Fig. 3b, we report the experimental momentum maps for both as-deposited (left half) and annealed (right half) porphyrin film, measured at the BEs marked in Fig. 3a. Whereas the sharp states close to the $\Gamma$ point are associated with the Cu(100) sp-bands, in all momentum maps the pattern related to the molecular overlayer is visible around $|\mathbf{k}| \approx 1.5$ Å$^{-1}$, as described in ref. [29]. These molecular features are associated with the HOMO−1 (BE: 1.7 eV), the degenerate LUMO/LUMO+1 (BE: 1.0 eV) and LUMO+3 (BE: 0.4 eV) of the NiTPP molecule, respectively. Upon annealing up to 620 K, we do not observe any remarkable change in these molecular patterns. The increased intensity of the LUMO+3 peak in the valence band spectra (see Fig. 3a) after annealing to 620 K is consistent with an increase of the charge transfer from the substrate to the molecule.

PT has already been proven to be a suitable tool for identifying possible intermediate reaction products, resulting from a stepwise dehydrogenation followed by a C–Cu bond formation, as demonstrated in the case of the thermally induced reaction of the transformation of dibromo-bianthracene into nano-graphene.[44] In our case, however, the one-to-one match between the momentum patterns of the as-deposited and the annealed molecular overlayers clearly shows that neither a cyclodehydrogenation at the macrocycle rim followed by a ring closure reaction nor a C–Cu bond formation takes place upon annealing, as otherwise the molecular symmetry would have been affected and the appearance in the reciprocal space of the corresponding molecular orbitals would have changed.

To confirm our interpretation, we performed DFT calculations for dehydrogenated NiTPP (dh-NiTPP) where the phenyl rings are fused with the macrocycle leading to benzotetracyclopenta[*at*,*ef*,*jk*,*op*]porphyrin, which is depicted in the inset in Fig. 3c, bottom. It should be noted that such reaction product, as resulting from the NiTPP dehydrogenation, is the only possible form that is compatible with the fourfold symmetry observed in STM imaging.

The calculated momentum distributions of the frontier molecular orbitals of dh-NiTPP are reported in Fig. 3c (top, red color scale) and compared to those of the pristine NiTPP (bottom, blue color scale). The resulting momentum patterns of dh-NiTPP, together with its overall electronic structure, significantly differ from the one of NiTPP and they do not resemble the measured maps in Fig. 3b. As an example, the LUMO+3 changes its π character becoming a σ orbital, and at the same time, the HOMO–LUMO gap decreases due to the delocalization of the π electrons over the whole molecule. Thus, by means of PT, we can rule out the ring-closing reaction between the phenyls and the macrocycle, because it would lead to a completely different appearance of the molecular features in *k*-space, which do not match the experimental maps after annealing. Further details of the energy level alignment of the gas-phase dh-NiTPP and the corresponding momentum distribution of all the frontier MO can be found in Section 4 of the ESI.†

Angle-dependent NEXAFS measurements can provide valuable information on both electronic structure and orientation with respect to the copper surface plane of a specific moiety. In Fig. 4a, we compare the NEXAFS spectra acquired across the N K-edge with p and s linear polarized light, for the as-deposited (bottom) and annealed (top) NiTPP film up to 490 K. We observe that, upon annealing, the NEXAFS dichroism does not change, and, at the same time, the π-symmetry resonances at 398.7 and 401.5 eV witness an overall intensity decrease.

These two resonances are mainly related to the transition of N 1s electrons to the degenerate LUMO/LUMO+1 and LUMO+3 orbitals and already in the as-deposited film they are strongly quenched with respect to their counterparts in a NiTPP multilayer or on oxygen-passivated copper[60] (used as a reference for weakly interacting molecules), because of the strong charge transfer taking place at the interface.[27] While no changes in dichroism indicate that the macrocycle remains both flat and parallel to the surface, the intensity decrease of these resonances points towards an overall enhancement of the charge transfer, from the substrate to the molecular film, in agreement with the trend in the VB spectra shown above and with the small core level shift to lower binding energy (<−0.1 eV) of the N 1s XPS observed upon

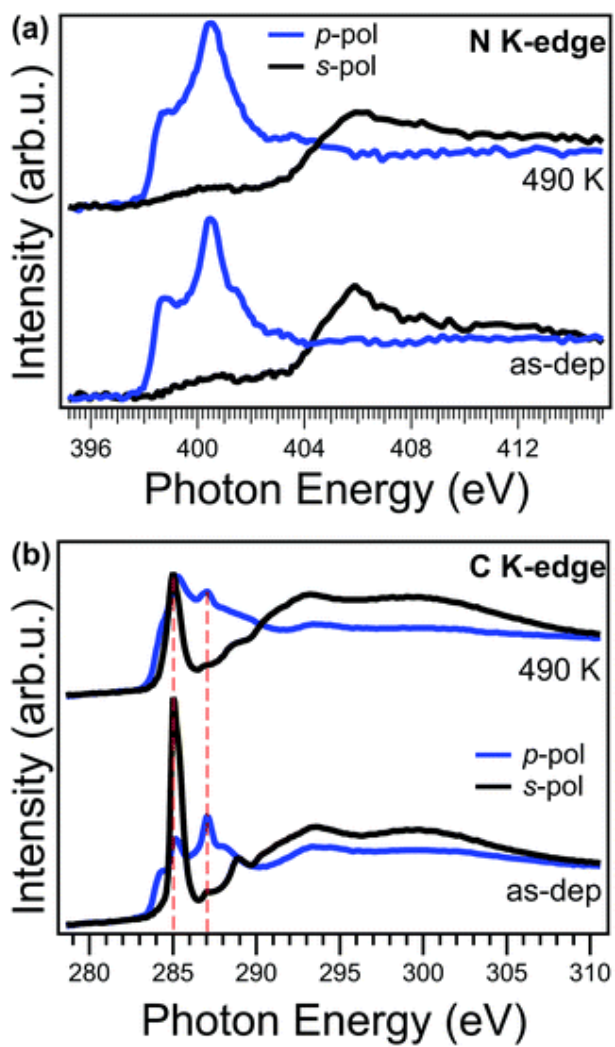

annealing to 420 K.

**Fig. 4** NEXAFS measurements of the NiTPP/Cu(100) interface. The spectra are acquired at the nitrogen (a) and carbon (b) K-edges for the as-deposited film (bottom panels) and after annealing up to 490 K (top panel).

In contrast to the N K-edge, the NEXAFS measurements at the C K-edge, reported in Fig. 4b, show, after the annealing, a change of the linear dichroism in the spectral feature at 285.1 eV mainly associated with the π* resonance of phenyls in NEXAFS C K-edge spectrum.[61] In fact, by looking at the intensity ratio of these resonance in the two polarizations, it is possible to estimate the average angle between the direction of the π* orbitals of the phenyl moiety and the surface normal (hereinafter tilt-off angle). While for the as-deposited NiTPP film the tilt-off angle of the phenyl results to be 72° ± 5°, it decreases up to 62° ± 5° after the thermal transition. In the present case, it has to be noticed that this tilt-off angle results from the convolution of both twisting and tilting of the phenyl ligand[29] and the two contributions cannot be disentangled solely from the NEXAFS data. Further details on the analysis of the NEXAFS spectra can be found in the ESI.† In any case, this small variation alone cannot fully explain the observed dimming of the phenyl features in STM images. While angle-dependent NEXAFS spectra allows us to determine the orientation of a specific molecular moiety with respect to the surface, the spectrum taken at the magic angle (54.7°) is more suitable for a visual appreciation of electronic structure changes in the moiety.[62] The spectrum at the magic angle can be simply obtained by the linear combination of the spectra in s- and p-pol (assuming a 100% linearly polarized light).[62] The spectra of the as-deposited and annealed NiTPP/Cu(100) system at the magic angle are reported in the ESI† (Fig. 8S). Upon annealing, the overall intensity decrease of the phenyl resonances at 285.1 and 289.0 eV, and the shift towards lower photon energies of the latter, indicates an increase of the charge transfer to the substituent,[63] which, combined with the observed large core level shift of the phenyl C 1s component, suggests also an electronic origin of the change of STM contrast. In further agreement with the XPS evidence, the NEXAFS C K-edge spectra also exclude the local formation of any C–Cu bond, which would give rise to a characteristic π* resonance in the s-pol spectrum at 288.1 eV, as was previously reported for *para*-phenylene polymerization on Cu.[31]

Summarizing the topographic and X-ray spectroscopy data, we conclude that the Ni(I)TPP monolayer phase undergoes an irreversible thermally-activated transformation resulting in an increase of the charge transfer into the phenyl substituents and a change of their tilt angle, while the electronic structure of the macrocycle and its adsorption geometry are preserved.

Infrared-visible sum-frequency generation (IR-Vis SFG) spectroscopy measurements endorse the conformational change model depicted above. In Fig. 5, we report the vibronic spectra measured in the 1240–3180 cm$^{-1}$ frequency region of the as-deposited monolayer at room temperature (top) and after stepwise annealing at progressively higher temperature values (from top to bottom). The NiTPP/Cu(100) system is characterized by a number of specific features that can be quantitatively determined by proper deconvolution of the experimental data according to the procedure described in the Methods section. The best results of the fitting procedure are shown in the figure (black line and color-filled profiles). The assignment to the respective vibrational modes can be obtained with reference to literature data (see Table S1, ESI† for quantitative details).[64–67]

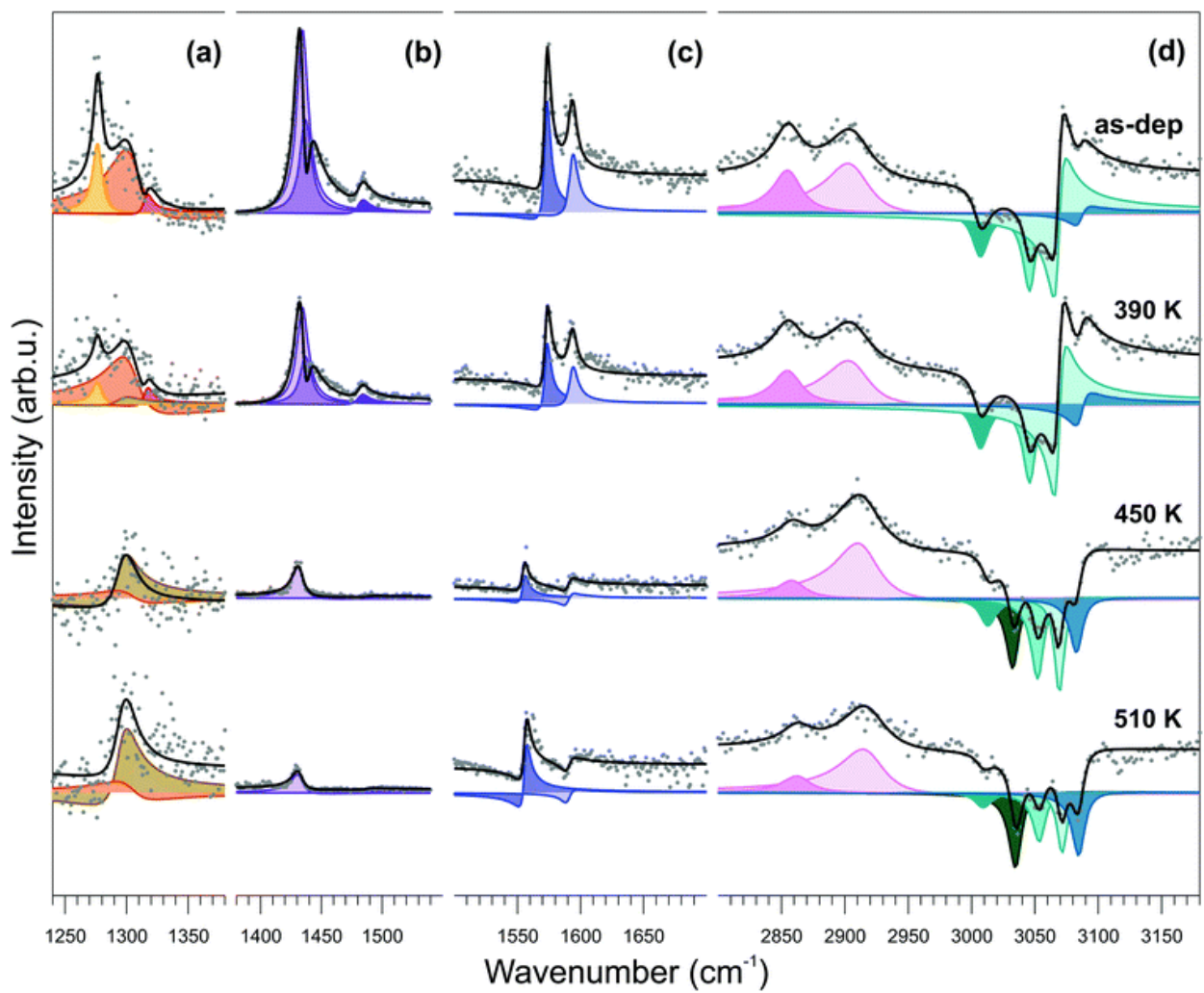

**Fig. 5** Temperature dependent IR-Vis SFG spectra of 1 ML NiTPP/Cu(100). Spectra were acquired at room temperature after annealing the system to progressively higher temperatures (from top to bottom). Data were renormalized and split into four different energy windows (a–d) for best visualization. The normalized SFG signal (grey dots) is shown together with the best fit (black lines) and the deconvolution (color filled offset profiles), according to the model lineshape described in the text. Data rescaling factors are (a) 1, (b) 0.15, (c) 0.5, (d) 0.2.

The temperature-dependent evolution of the vibronic spectra in the 300–510 K range indicates substantial stability of the molecule for both the macrocycle and the phenyl moieties. An overall increase of planarity of the molecule is suggested by the general reduction of resonant amplitudes at increasing temperatures. More specifically, from the quantitative evolution of the resonant amplitudes, line positions, and relative phases (see ESI,† Section 6), a number of conclusions can be extracted from the data. First, the feature growing with the temperature at 1296 cm$^{-1}$ (Fig. 5a) at the expense of the resonances at 1277, 1304, and 1316 cm$^{-1}$ is associated with the orientational rearrangement of the out-of-plane modes of the phenyl rings. Second, the strong phase changes in the C–H high energy region (*d*), accompanied by the growth of a new resonance at 3034 cm$^{-1}$, indicate both reorientation of the phenyl rings and increase of their interaction with the underlying surface. Finally, the progressively increasing non-resonant amplitude (shown in Fig. 9S of the ESI†) indicates modifications of the surface electronic charge density within 2.3 eV from the Fermi level, due to a change of the molecule–surface interaction. Altogether, these observations exclude the fusion of the phenyl moieties with adjacent molecules or with the macrocycle and rather point in the direction of substantial temperature stability of the NiTPP monolayer, at least up to 510 K. The observed downward bending of the phenyl groups is associated with the opening of a new vibrational channel and corresponding increase of the phenyl interaction with the copper substrate, which finally determines also the large core level shift of the C 1s phenyl component, as measured by XPS.

What is the ultimate origin of the mechanism driving the transformation? All the experimental data, consistently indicate that the molecular changes are confined to the phenyl moiety, while the macrocycle remains rigidly anchored to the substrate. We can obtain a rough estimate of ∼100 kJ mol$^{-1}$ (∼1 eV) for the activation barrier of the phenyl transition, from the plot of the temperature-dependent intensity of the corresponding C 1s peak at 284.4 eV (Fig. 6S, in the ESI†). This value is much smaller than typical barriers for flattening by cyclodehydrogenation, which vary with the degree of flexibility of the macrocycle and the strength of intermolecular interaction. On low reactive substrates, such as Ag, where the macrocycle flexibility is larger than on Cu, the cyclodehydrogenation takes place well beyond 500 K[22,24,45] and even higher for a metallated TPP (due to the increased rigidity of the macrocycle).[25,68] On Cu(111), where the overall charge transfer at the interface is lower than on Cu(100), as evaluated from the intensity of resonances at 398.7 eV in the NEXAFS N K-edge,[69] the cyclodehydrogenation is completed at 570–590 K: the higher the coverage, the higher the temperature, because of the additional stability of molecular islands due to intermolecular phenyl–phenyl interaction.[23]

An attempt to estimate the energy cost of simply twisting the phenyls can be done with literature data. In the case of a perfectly rigid and flat macrocycle, closely resembling our case, first principle calculations for isolated M-TPP obtain an energy cost <0.1 eV for a 20° twisting around a mean value of 60°,[70] which is expected to increase in case of additionally upward bent phenyls, due to the presence of the substrate. A larger energy barrier certainly comes from the intermolecular phenyl–phenyl interaction. This can be estimated from the benzene–benzene electrostatic attraction, which is in the range of 0.1 eV per pair for an edge to face geometry (T-shape),[71] similar to the present phenyl configuration. As a consequence, only considering intramolecular barriers and direct phenyl–phenyl intermolecular interaction, a minimum energy barrier of 0.5 eV can be estimated for the present conformational change in a free-standing layer. Moreover, the charge transferred from the substrate to the molecule may easily increase the latter energy barrier. In this regard, a change of the local surface relaxation/reconstruction might be the underlying mechanism that stabilizes the new configuration. In fact, massive reshaping of the Cu(111) surface upon deposition of metallated phthalocyanines takes place already at room temperature.[72] More generally, the surfaces of copper are subject to extensive patterning by extended aromatic complexes that can capture or extract the adatoms beneath.[73,74] Even if we could not detect the formation of adatoms, our C 1s XPS and IR-Vis SFG results agree with a stronger coupling between the downward tilted phenyls and the surface Cu atoms.

Finally, the molecular pinning due to the molecule–substrate interaction, even strengthened by the transition, prevents further flattening of the phenyls up to the molecular decomposition. For comparison, flattening by cyclodehydrogenation on the rutile $TiO_2$(110) surface takes place only at very high temperature (∼700 K) due to the strong chemical bonding of the tetrapyrrolic pocket to the substrate.[48] We can conclude that the extent of the charge transfer at the molecule–surface interface is a key factor for the activation of cyclodehydrogenation reactions on metal substrates.

## Conclusions

STM indicates the occurrence of a gradual and irreversible modification of the phenyl terminations upon thermal annealing at 420 K and beyond. This, however, does not modify the overall molecular orientation and spacing, as well as the phase symmetry. At the same time, the so-obtained molecular conformation preserves the active Ni(I) oxidation state of the center up to the limit of thermal decomposition (620 K), as proven by XPS of the Ni 2p and C 1s core levels. By means of photoemission tomography, we could state that Ni(I)TPP molecules adsorbed on Cu(100) preserve the electronic structure and symmetry of the as-deposited film, excluding any possible chemical modification of the heteroaromatic molecular configuration (*e.g.*, partial cyclodehydrogenation of the phenyls).

From angular dependence of linearly polarized NEXAFS measurements, we could determine the thermally activated transition to be associated with a conformational change of the orientation of the phenyls. An investigation of the vibrational states with IR-Vis SFG spectroscopy shows that upon annealing a strong interaction of the phenyls with the substrate develops upon annealing, in full agreement with the observed shift towards lower binding energy of the phenyl component in the XPS of the C 1s core level.

Our data demonstrate that the strong molecule–substrate interaction at the basis of the macrocycle reduction results in an effective pinning of the molecule to its adsorption site, thus preventing any molecular displacement in the overlayer lattice, which would eventually allow the normally observed molecular flattening by cyclodehydrogenation, *e.g.*, the modification of the molecular heteroaromatic structure. The strong thermal stability of the Ni(I)TPP network supported by copper paves the way to the application of this reactive interface in the field of heterogeneous catalysis, gas sensing and surface magnetochemistry, where temperatures up to 620 K can be used to regenerate the pristine properties of the active molecular porphyrin layer.

## Author contribution

H. S. and I. C. carried out the photoemission experiments and the consequent data analysis with support from G. Z., V. F., M. J., E. V., M. S., A. V., A. C. and L. F. P. P. performed the DFT simulations. A. S. carried out the STM measurements and M. S., S. M., M. C. and E. V. performed IR-Vis SFG experiments and the consequent data analysis. G. Z., V. F., H. S. and I. C., with the assistance of M. C., L. F., A. S. and E. V., drafted the manuscript, which was discussed together with C. M. S. and G. C. All authors discussed the results and reviewed the manuscript.

## Data availability

The authors declare that relevant data supporting the findings of this study are available on request.

## Conflicts of interest

There are no conflicts to declare.

## Acknowledgements

This work has partially received funding from the EU-H2020 research and innovation programme under grant agreement no. 654360 having benefitted from the access provided by CNR-IOM in Trieste (Italy) within the framework of the NFFA Europe Transnational Access Activity. M. C., G. Z. and H. S. acknowledge funding from the European Research Council (ERC) under the European Union's Horizon 2020 research and innovation programme (Grant Agreement No. 725767—hyControl). Support from Italian MIUR under project PRIN-2017KFY7XF is also acknowledged. P. P. acknowledges support from the Austrian Science Fund (FWF) project I3731. The authors gratefully thank Cristina Africh, IOM-CNR, Italy, for fruitful discussion.

## Notes and references

## Footnotes

1. † Electronic supplementary information (ESI) available. See DOI: 10.1039/d0tc00946f
2. ‡ Present address: Istituto di Struttura della Materia-CNR (ISM-CNR), Trieste, 34149, Italy.